\documentclass[12pt]{iopart}
\usepackage{iopams}  
\usepackage{setstack}
\usepackage{times,graphicx,xcolor}
\usepackage{delarray,fancybox}
\usepackage{mathdots}
\newcommand{\eps}{\varepsilon}

\newcommand{\sbr}[1]{\left[#1\right]}
\newcommand{\cbr}[1]{\left\{#1\right\}}
\newcommand{\de}{\partial}
\newcommand{\nn}{\nonumber \\}

\newcommand{\flu}[1]{\delta #1 }

\begin{document}

\title[Scaling regimes of 2d turbulence]{Scaling regimes of 2d turbulence with power law stirring:
theories versus numerical experiments}
%2d turbulence:RG vs numerical experiments]{2d turbulence: 
%renormalization group theory versus numerical experiments.
%Scaling regimes of 2d turbulence with power law stirring}

\author{A.~Mazzino}
\address{Department of  Physics, University of Genova, INFN and CNISM, 
via Dodecaneso 33, 16146 Genova, Italy.}
\ead{andrea.mazzino@unige.it}

\author{P.~Muratore-Ginanneschi}
\address{Department of Mathematics and Statistics, University of Helsinki PL 68, 
00014 Helsinki, Finland.}
\ead{paolo.muratore-ginanneschi@helsinki.fi}

\author{S.~Musacchio}
\address{Department of Physics, INFN and CNISM, University of Torino, via P.Giuria 1, 
10125 Torino, Italy.}
\ead{stefano.musacchio@gmail.com}

\pacs{47.27.Te, 47.27.$-$i, 05.10.Cc00.00, 20.00, 42.10}

\begin{abstract}
We inquire the statistical properties of the pair formed by   
the Navier--Stokes equation for an incompressible velocity field 
and the advection-diffusion equation for a scalar field transported 
in the same flow in \emph{two dimensions} ($2d$). The system is in a regime of 
fully developed turbulence stirred by forcing fields with Gaussian statistics, 
white-noise in time and self-similar in space. 
%Fluctuations of the passive scalar field are sustained by an analogous source
%term in the advection-diffusion equation. 
%This setting is motivated by the possibility to carry out  
In this setting and if the stirring is concentrated at small spatial scales as
if due to thermal fluctuations, it is possible to carry out a first-principle 
ultra-violet renormalization group analysis of the scaling behavior of the model. 
%At variance with the three dimensional case, 
%The 
Kraichnan's phenomenological theory 
%of Kraichnan, Leith and Batchelor 
of two dimensional turbulence upholds the existence of an inertial range characterized 
by inverse energy transfer at scales larger than the stirring one. 
%A-priori, when the H\"older exponents
%of the forcing sustaining the velocity and the scalar fields  
%describe stirring by thermal fluctuations, renormalization group 
%analysis applies in a parametric domain
%where, according to Kraichnan's phenomenological theory,
%an inverse cascade should occur.
For our model Kraichnan's theory, however, implies scaling predictions 
radically discordant from the renormalization group results.
We perform accurate numerical experiments to assess the actual statistical 
properties of $2d$-turbulence with power-law stirring. Our results clearly indicate 
that an adapted version of Kraichnan's theory is consistent with 
the observed phenomenology. We also provide some theoretical scenarios to account 
for the discrepancy between renormalization group analysis and the observed
phenomenology. 
\end{abstract}

\section{Introduction}

Two-dimensional ($2d$) turbulence is interesting for several reasons. 
In laboratory experiments $2d$ turbulence has been realized and studied 
with electromagnetically driven liquid metals 
\cite{PaTa98,PaJuTa99,BoCeEsMu05} and thin soap films
\cite{RiVoEc98,RiVoEc99,RiDaChEc03}.
In geophysical flows, vertical confinement suggests the possibility to
describe the mesoscale dynamics of atmosphere and oceans in terms of
two-dimensional fluid models \cite{Ch47,Ea49}.
Indeed, observational data such as the Nastrom-Gage spectrum \cite{NaGa85,NaGaJa84},
studies based on the MOZAIC database \cite{MOZAIC} and on the EOLE Lagrangian 
balloons in the low stratosphere \cite{LaAuLeVu04} support the existence of a 
mesoscale $-5/3$ power-law energy spectrum which may be the consequence 
of a two-dimensional inverse cascade. 
Although recent studies \cite{Li99b,LiCh01} suggest the occurrence of a fairly 
more complex physical phenomenology (see e.g. \cite{TuOr03,XiPuFaSh08}), the
$2d$-approximation remains an important benchmark for understanding the atmospheric 
physics at synoptics and planetary scales \cite{Li07} as well as in other 
geophysical contexts (see \cite{Ta02,IUTAM98} and references therein). 
For example, analysis of spectral kinetic energy fluxes in satellite altimeter data 
provides strong evidences of the occurrence of an inverse energy cascade 
in the ocean \cite{ArFlSc06}. 
From the point of view of statistical mechanics,  
$2d$-turbulence is a prototype of  non-equilibrium systems
whose steady state is not described by Boltzmann statistics.

At variance with its three-dimensional counterpart for which the total kinetic energy is
the unique inviscid invariant, the two-dimensional Navier--Stokes equation
preserves also the total enstrophy (in the absence of forcing and dissipation) \cite{Frisch}.
Enstrophy conservation is a key ingredient for 
the proof \cite{La69} of the existence and uniqueness of the solution of 
the Cauchy problem for the $2d$ Navier--Stokes equation with deterministic forcing. 
Very recently, this result has been extended to stochastic stirring. 
In particular it was shown \cite{KuSh00, BrLeKu00, EMaSi01, BrLeKu01, BrLeKu02, Ma02} that  
the solution is a Markov process exponentially mixing in time and ergodic with a unique 
invariant (steady state) measure even when the forcing acts only on two Fourier modes 
\cite{HaMa04}.
 
A \emph{phenomenological} theory proposed by Kraichnan \cite{Kr67} and further extended by 
Batchelor \cite{Ba69} and Leith \cite{Le68} (see also \cite{KrMo80,Be99a,Be99b,Li99b,GkTu07}) 
predicts the presence of a double cascade mechanism governing the transfer of energy 
and enstrophy in the limit of infinite inertial range. 
Accordingly, an \emph{inverse energy cascade} with spectrum 
characterized by a scaling exponent $-5/3$ appears for values of the 
wave-number $p$ smaller than $p_{f}$, the typical  forcing wave-number.
For wave-numbers larger than $p_{f}$ a \emph{direct enstrophy cascade} should occur. 
In this regime the energy spectrum should have a power-law exponent equal to $-3$
plus possible logarithmic corrections hypothesized in \cite{Kr71} to ensure 
constancy of the enstrophy transfer rate. 
Very strong laboratory experiments reviewed in \cite{Ta02,KeGo02}, and
numerical experiments (see e.g. \cite{Bo06} and references therein) corroborate
Kraichnan's theory. 
A long standing hypothesis \cite{Po92}, 
%recently corroborated by numerical experiments 
which has recently found support through numerical experiments
\cite{BeBoCeFa06,BeBoCeFa06b}, also surmises the existence of a conformal
invariance underlying the inverse energy cascade of $2d$ turbulence. 

Despite these successes, a \emph{first principle} derivation of the statistical properties 
of $2d$ turbulence is still missing. An attempt in this direction has recently been undertaken 
\cite{Ho98,HnHoHoSe99,HoKaKo02,AdHoKoVa05} by inquiring the scaling properties 
of the velocity field and of the transported scalar field (passive scalar) 
when they are sustained by a random Gaussian forcing with self-similar
spatial statistics. The H\"older exponent $\varepsilon$ of the forcing correlation provides
an order parameter interpolating between small-scale thermal stirring and large-scale
stirring.

In three (and more) dimensions ultra-violet renormalization group 
analysis \cite{FoNeSt77,DeMa79} of this model yields the result
 $1-4\varepsilon/3$ for the scaling exponent of the kinetic energy spectrum 
holding to \emph{all orders} in a perturbative expansion in powers of $\varepsilon$.
Kolmogorov scaling \cite{Frisch} is recovered when the energy input becomes 
dominated by its infra-red components at $\varepsilon$ equal two. 
The results of \cite{FoNeSt77,DeMa79} are coherent with physical intuition 
because only the case $\varepsilon$ equal two is a model of fully developed turbulence. 
 Recent numerical simulations validate within available resolution such picture 
\cite{SaMaPa98,BiLaTo04}. 

The extension of renormalization group analysis 
to the $2d$ case is instead not straightforward and was only achieved in \cite{Ho98}. 
Although the prediction for the kinetic energy scaling exponent is the same 
as in the three dimensional case, the result cannot be easily reconciled with the 
phenomenological intuition based on Kraichnan's theory. The latter suggests 
the onset of an inverse energy cascade already at $\varepsilon$ equal zero when the energy and 
enstrophy input is dominated by the ultra-violet degrees of freedom. 

The purpose of the present work is to shed light on the apparent contradiction 
between the %two theories
phenomenological and the renormalization group theory. 
In doing so, we extend and complete results presented in a previous letter 
\cite{MaMGMu07}.
In particular, in section~\ref{sec:model} we illustrate 
the details and interpretation of the model. In section~\ref{sec:rg} 
we summarize the results of the renormalization group analysis found in 
\cite{Ho98,HnHoHoSe99,HoKaKo02,AdHoKoVa05}. 
In section~\ref{sec:Kraichnan} we draw on \cite{Be99a,Be99b} to solve the
K\'arm\'an-Howarth-Monin equation under the hypotheses of Kraichnan's theory. 
In doing so, we focus first our attention on the parametric range where a 
direct comparison with the renormalization group theory is possible. We also 
shortly discuss the predictions of the phenomenological theory for the general 
case, as a benchmark for our numerical experiments. 
In section~\ref{sec:numerics} we compare the renormalization group predictions 
with the outcomes of a direct numerical integration of the model equations.
There we give clear evidences that in all parametric ranges only the phenomenological
theory \emph{\`a la} Kraichnan is able to describe the observed behavior of the 
stochastic flow. 
Finally, in section~\ref{sec:discussion} we discuss possible mechanisms underlying 
the discrepancy between the renormalization group predictions 
 and the observed scaling behavior of the model.

\section{The model}
\label{sec:model}

We consider the $2d$ Navier--Stokes equation governing the
evolution of the velocity $\boldsymbol{v}$ of an incompressible Newtonian fluid
\begin{eqnarray}
&& (\partial_{t}+\boldsymbol{v}\cdot\boldsymbol{\de}_{\boldsymbol{x}})\,\boldsymbol{v}
=\nu\,\partial_{\boldsymbol{x}}^{2}\,\boldsymbol{v}-\boldsymbol{\partial_{x}}P 
-\frac{\boldsymbol{v}}{\tau}+
\boldsymbol{f}
\label{model:ns}
\\
&&\boldsymbol{\partial}_{\boldsymbol{x}}\cdot \boldsymbol{v}=
\boldsymbol{\partial}_{\boldsymbol{x}}\cdot \boldsymbol{f}=0
\label{model:incompressibility}
\end{eqnarray} 
and the forced advection--diffusion equation for a scalar field $\theta$.  
\begin{eqnarray}
 (\partial_{t}+\boldsymbol{v}\cdot\boldsymbol{\partial_{x}})\,\theta=
\kappa\,\partial_{x}^{2}\,\theta+g
\label{model:ps}
\end{eqnarray}
where $\nu$ is the kinematic viscosity and $\kappa$ the diffusivity of the scalar field. 
The Ekman friction term $-\frac{\boldsymbol{v}}{\tau}$ included in Eq.(\ref{model:ns}) 
ensures that a steady-state is attained by damping kinetic energy transfer towards larger
and larger scales \cite{Fj53}. Both equations (\ref{model:ns}) and (\ref{model:ps}) 
are sustained by stochastic forcing fields, 
respectively $\boldsymbol{f}$ and $g$,  with Gaussian 
statistics such that 
\begin{eqnarray}
\prec \boldsymbol{f}(\boldsymbol{x},t) \succ=\prec g(\boldsymbol{x},t) \succ=0
\label{model:zeroav}
\end{eqnarray}
and
\begin{eqnarray}
\label{model:nsforce}
\prec f^{\alpha}(\boldsymbol{x},t)f^{\beta}(\boldsymbol{y},s)\succ=
\delta(t-s)\,F^{\alpha\,\beta}(\boldsymbol{x}-\boldsymbol{y},m_{f},M_{f})\,,\hspace{0.5cm}
\alpha\,,\beta=1,2
\end{eqnarray}
\begin{eqnarray}
\label{model:psforce}
\prec g(\boldsymbol{x},t)g(\boldsymbol{y},s)\succ=
\delta(t-s)\,G(\boldsymbol{x}-\boldsymbol{y},m_{g},M_{g})
\end{eqnarray}
Whilst time decorrelation in (\ref{model:nsforce}), (\ref{model:psforce}) 
is meant to preserve Galilean invariance of the statistics (in the absence of Ekman friction), 
the spatial part of the forcing correlations is chosen to be isotropic and 
self-similar in a wave-number range between well separated 
infra-red $m_{f}$, $m_{g}$ and ultra-violet $M_{f}$, $M_{g}$ cut-offs. 
Specifically
\begin{eqnarray}
\label{model:F}
\hspace{-1.0cm}
F^{\alpha\,\beta}(\boldsymbol{x},m_{f},M_{f})=F_{o}\int\,\frac{d^{2}p}{(2\,\pi)^{2}}
\frac{e^{\imath\,\boldsymbol{p}\cdot\boldsymbol{x}}}{p^{2\,\varepsilon-2}}\,
\Pi^{\alpha\,\beta}(\boldsymbol{\hat{p}})\,
\chi_{f}\left(\frac{m_{f}^{2}}{p^{2}},\frac{p^{2}}{M_{f}^{2}}\right)\,,
\hspace{0.5cm} \frac{m_{f}^{2}}{M_{f}^{2}}\,\ll\,1
\end{eqnarray}
with $\Pi^{\alpha\,\beta}$ the transversal projector in Fourier space, $\varepsilon$ the
H\"older exponent and 
\begin{eqnarray}
\label{model:G}
G(\boldsymbol{x},m_{g},M_{g})=G_{o}\int\frac{d^{2}p}{(2\,\pi)^{2}}\,
\frac{e^{\imath\,\boldsymbol{p}\cdot\boldsymbol{x}}}{p^{2\,h-2}}
\,\chi_{g}\left(\frac{m_{g}^{2}}{p^{2}},\frac{p^{2}}{M_{g}^{2}}\right)\,,
\hspace{0.5cm} \frac{m_{g}}{M_{g}}\,\ll\,1
\end{eqnarray}
with H\"older exponent $h$. The explicit form of the isotropic cut-off 
functions $\chi_{f}$ and $\chi_{g}$ is unimportant  
as far as they remain approximately 
constant for any wave number $\boldsymbol{p}$ in the scaling range $\label{model:scaling}
m_{f}\,,m_{g}\,\ll\,p\,\ll\,M_{f}\,,M_{g} $. 
It is not restrictive to choose $\chi_{f}$, $\chi_{g}$ normalized to the
unity in the origin. Note that at $\varepsilon$, $h$ equal zero the traces of
(\ref{model:F}) and (\ref{model:G}) are proportional to the Laplacian of
a Dirac $\delta$-function centered at the origin. More generally, the 
H\"older exponents $\varepsilon$, $h$ determine the spectral composition 
of the energy injection. For the velocity field one finds
\begin{eqnarray}
\label{model:Einjection}
\hspace{-2.2cm}I_{\mathcal{E}}(m_{f},M_{f})=F_{o}\int\frac{d^{2}p}{(2\,\pi)^{2}}
\,p^{2-2\,\varepsilon}\,\chi_{f}\left(\frac{m_{f}^{2}}{p^{2}}\,,\frac{p^{2}}{M_{f}^{2}}\right)
\propto\left\{
\begin{array}{ll}
%\frac{M_{f}^{4-2\,\varepsilon}\,F_{o}}{4-2\,\varepsilon}
M_{f}^{4-2\,\varepsilon}\,I_{\mathcal{E}}(0,1)\,,
\hspace{0.2cm} & \hspace{0.2cm} 0\,\leq\,\varepsilon\,<\,2
\\[0.5cm]
%\frac{m_{f}^{4-2\,\varepsilon}\,F_{o}}{2\,\varepsilon-2} 
m_{f}^{4-2\,\varepsilon}\,I_{\mathcal{E}}(1,0)\,,
\hspace{0.2cm} & \hspace{0.2cm} \varepsilon\,>\,2
\end{array}
\right.
\end{eqnarray}
Similarly, the injection for the scalar field is dominated by 
wave-numbers around the ultra-violet (infra-red) cut-off for any $h\,<\,2$ 
($h\,>\,2$). For the purposes of the present analysis it is worth recalling
the vorticity representation of Navier--Stokes equation in $2\,d$:
\begin{eqnarray}
\label{model:vorticity}
\left(\partial_{t}+\boldsymbol{v}\cdot\partial_{\boldsymbol{x}}\right)\omega
=\nu\,\partial_{\boldsymbol{x}}^{2}\omega-\frac{\omega}{\tau}+f_{\omega}
\end{eqnarray}
where
\begin{eqnarray}
\begin{array}{lll}
\omega=\epsilon_{\alpha\,\beta}\partial_{x_{\alpha}}v^{\beta}\hspace{0.5cm}& \&\hspace{0.5cm}&
f_{\omega}=\epsilon_{\alpha\,\beta}\partial_{x^{\alpha}}f^{\beta}
\\[0.2cm]
\epsilon_{1\,2}=- \epsilon_{2\,1}=1 
\hspace{0.5cm}& \&\hspace{0.5cm}&\epsilon_{1\,1}=\epsilon_{2\,2}=0
\end{array}
%\omega=\epsilon_{\alpha\,\beta}\partial_{x_{\alpha}}v^{\beta}
%\hspace{0.5cm} \&\hspace{0.5cm}
%f_{\omega}=\epsilon_{\alpha\,\beta}\partial_{x^{\alpha}}f^{\beta}
%\hspace{0.5cm} \&\hspace{0.5cm}
%\epsilon_{1\,2}=- \epsilon_{2\,1}=1 
%\hspace{0.5cm} \&\hspace{0.5cm}\epsilon_{1\,1}=\epsilon_{2\,2}=0
\end{eqnarray}
Equation (\ref{model:vorticity}) implies the conservation of the total 
enstrophy $\mathcal{Z}$
\begin{eqnarray}
\label{model:enstrophy}
\mathcal{Z}=\int d^{2}x\,\, \prec \omega(\boldsymbol{x},t)\omega(\boldsymbol{0},t) \succ
=-\int d^{2}x\,\, \partial_{\boldsymbol{x}}^{2}
\prec \boldsymbol{v}(\boldsymbol{x},t)\cdot\boldsymbol{v}(\boldsymbol{0},t) \succ
\end{eqnarray}
whenever the right hand side of (\ref{model:vorticity}) is negligible. 
For power-law forcing, the enstrophy injection is also controlled by the 
H\"older exponent $\varepsilon$
\begin{eqnarray}
\label{model:Zinjection}
\hspace{-2.3cm}
I_{\mathcal{Z}}(m_{f}\,,M_{f})=F_{o}\int\frac{d^{2}p}{(2\,\pi)^{2}}
p^{4-2\,\varepsilon}\,\chi_{f}\left(\frac{m_{f}^{2}}{p^{2}}\,,\frac{p^{2}}{M_{f}^{2}}\right)
\propto\left\{
\begin{array}{ll}
%\frac{M_{f}^{6-2\,\varepsilon}\,F_{o}}{6-2\,\varepsilon}\,,
M_{f}^{6-2\,\varepsilon}\,I_{\mathcal{Z}}(0,1)\,,
\hspace{0.2cm} & \hspace{0.2cm} 0\,\leq\,\varepsilon\,<\,3
\\[0.5cm]
%\frac{m_{f}^{6-2\,\varepsilon}\,F_{o}}{2\,\varepsilon-6}\,, 
m_{f}^{6-2\,\varepsilon}\,\,I_{\mathcal{Z}}(1,0)\,,
\hspace{0.2cm} & \hspace{0.2cm} \varepsilon\,>\,3
\end{array}
\right.
\end{eqnarray}
The relations (\ref{model:Einjection}), (\ref{model:Zinjection}) show 
that the energy and enstrophy injections are simultaneously concentrated
in the ultra-violet and in the infra-red only for 
$\varepsilon\,<\,2$ and $\varepsilon\,>\,3$, respectively. 

\section{Summary of the renormalization group analysis results}
\label{sec:rg}

Let $d_{A}$ denote the scaling dimension of a physical
quantity $A$. Crudely matching canonical dimensions in (\ref{model:ns}) 
and (\ref{model:ps}) suggests, irrespectively of the spatial dimension, 
the existence of two scaling ranges : a dissipative one for wave-numbers
such that non-linear effect are negligible
\begin{eqnarray}
\label{rg:dissipative}
d_{t}\,=\,2\,d_{x}
\hspace{0.3cm} \& \hspace{0.3cm} 
d_{v}\,=\,-\,d_{x}(1-\varepsilon)
\hspace{0.3cm} \& \hspace{0.3cm} 
d_{\theta}\,=\,-\,d_{x}(1-h)
\end{eqnarray} 
and an inertial range corresponding to the requirement of Galilean 
invariance imposed by matching the two terms of the material derivative
 $D_{t}:=\partial_{t}+\boldsymbol{v}\cdot\boldsymbol{\partial}_{\boldsymbol{x}}$ with the forcing
\begin{eqnarray}
\label{rg:inertial}
\hspace{-1.3cm}
d_{t}\,=\,d_{x}\,\left(2-\frac{2\,\varepsilon}{3}\right)
\hspace{0.3cm} \& \hspace{0.3cm} 
d_{v}\,=\,-\,d_{x}\left(1-\frac{2\,\varepsilon}{3}\right)
\hspace{0.3cm} \& \hspace{0.3cm} 
d_{\theta}\,=\,-\,d_{x}\left(1-h+\frac{\varepsilon}{2}\right)
\end{eqnarray}  
For $\varepsilon\,,h$ equal zero the scaling dimensions 
(\ref{rg:dissipative}), (\ref{rg:inertial}) coalesce. 
Physically, the coalescence point corresponds to stirring by
thermal noise. Coalescence hints at the existence of a marginal
case in renormalization group sense. This means (see e.g. \cite{Zinn}) that 
scaling dimensions at small but finite $\varepsilon\,,h$ may be obtained in a
Taylor series based at their values in the marginal case in analogy to 
equilibrium critical phenomena where marginality is defined
by an upper critical dimension specified e.g. by Ginzburg's 
criterion. A similar scenario seems to apply to (\ref{model:ns}) 
for $d\,>\,2$ and in the absence of large scale friction
($\tau$ set to infinity). In such a case, \cite{FoNeSt77,DeMa79,AdAnVa}   
fine-tuning the amplitude of the forcing correlation (\ref{model:F}) to
be $O(\varepsilon)$ yields an expansion in powers of $\varepsilon$ around 
a Gaussian theory specified 
by renormalized eddy diffusivities $\nu_{R}$ and $\kappa_{R}$.
Ultra-violet renormalization guarantees that the eddy diffusivities
are related to the molecular viscosities appearing in (\ref{model:ns})
and (\ref{model:ps}) by renormalization constants
\begin{eqnarray}
\label{rg:constants}
Z_{\nu}:=\frac{\nu}{\nu_{R}}\hspace{1.0cm} \& \hspace{1.0cm} Z_{\kappa}:=\frac{\kappa}{\kappa_{R}}
\end{eqnarray}
determined at any order in perturbation theory by subtracting all 
"resonant" terms diverging with the ultra-violet cut-offs $M_{f}$ and $M_{g}$. 
Technically, this is achieved by identifying the ultra-violet 
divergent part of the one-particle irreducible vertex associated to the 
response functions of the velocity and concentration fields \cite{AdAnVa}. 
The result is that, within all order accuracy in $\varepsilon$, all correlation functions 
of velocity and concentration fields sampled at well separated spatial points 
%do not admit any self-similarity breaking by the ultra-violet cut-offs. 
scale according to canonical dimensional predictions.
In particular, the scaling laws (\ref{rg:dissipative}) and (\ref{rg:inertial}) 
correspond to an ultra-violet and an infra-red stable fixed point 
of the renormalization group transformation respectively describing the dissipative 
and inertial ranges. Extending this analysis to the $2d$ case presents extra 
difficulties. Already in the absence of non-linearities, the correlation 
function is simultaneously logarithmically divergent both in the infra-red 
and in the ultra-violet. The Ekman term in (\ref{model:ns}) is then needed 
to decouple infra-red degrees of freedom. Once this is done, 
dimensional analysis shows that the renormalization constants (\ref{rg:constants}) 
are not sufficient to reabsorb all terms divergent with the ultra-violet 
cut-offs in the perturbative solution of (\ref{model:ns}). 
This is a serious difficulty because direct calculations \cite{AdHoKoVa05,Va04} 
hint %at the impossibility to construct ultra-violet renormalization 
that ultra-violet renormalization
group transformations using \emph{non-local} counter-terms may lead 
to mathematical inconsistencies. 
In other words, renormalization constants %can 
should only be associated to coupling constants of local interactions in real space. 
For (\ref{model:ns}) the molecular viscosity is the only coupling constant 
satisfying %the 
such requirement \cite{AdAnVa,Ho98,AdHoKoVa05}. 
The difficulty led to a controversy, summarized in \cite{Ho98}, 
about the very possibility to apply renormalization group methods to $2d$ turbulence. 
In \cite{Ho98} it is also argued that multiplicative ultra-violet renormalization
remains consistent to all orders in perturbation theory if the forcing correlation 
is modified to include a local (analytic) component
\begin{eqnarray}
\label{rg:replacement}
F^{\alpha\,\beta}(\boldsymbol{x},m_{f},M_{f})\,\to\,
F^{\alpha\,\beta}(\boldsymbol{x},m_{f},M_{f})+
F_{(local)}^{\alpha\,\beta}(\boldsymbol{x},m_{f},M_{f})
\end{eqnarray}
with
\begin{eqnarray}
\label{rg:localforcing}
F^{\alpha\,\beta}_{(local)}(\boldsymbol{x},m_{f},M_{f})\,:=
\,F_{o}^{(local)}\int\frac{d^{2}p}{(2\,\pi)^{2}}
\,e^{\imath \boldsymbol{p\cdot x}}
\Pi^{\alpha\,\beta}(\boldsymbol{\hat{p}})\,p^2\,
\chi_{f}\left(\frac{m_{f}^{2}}{p^{2}},\frac{p^{2}}{M_{f}^{2}}\right)
\end{eqnarray}
From the renormalization group point of view, the replacement is justified
by the observation that the resulting model generates the same relevant couplings
as the original and therefore should fall in the same universality class.
The merit of (\ref{rg:replacement}) is to provide through $a_{o}:=F_{o}^{(local)}/\nu^{3}$ 
the zeroth order of the extra renormalization constant
\begin{eqnarray}
\label{rg:extra}
Z_{a}:=\frac{a_{o}}{a}
\end{eqnarray}
needed to reabsorb all the remaining explicit dependence on $M_{f}$ 
in the perturbative expansion of correlation functions of the velocity field. 
In consequence, \cite{Ho98} predicts that the isotropic energy spectrum 
of the velocity field
\begin{eqnarray}
  \label{rg:Espectrum}
  E_{[\boldsymbol{v}]}(p):=\int\frac{d^{2}q}{(2\,\pi)^{2}}\,\delta(q-p)\int d^{2}x\,e^{\imath\,\boldsymbol{q\cdot x}}
\prec\boldsymbol{v}(\boldsymbol{x},t)\cdot\boldsymbol{v}(0,t)\succ
\end{eqnarray}
admits the expression
\begin{eqnarray}
\label{rg:energy}
E_{[\boldsymbol{v}]}(p)=\eps^{1/3}\,\left(\frac{F_{o}}{\nu^{3}}\right)^{2/3}\,\nu^{2}\,
p^{ 1-\frac{4\,\varepsilon}{3}}\,R\,\left(\varepsilon\,,\frac{m_{f}}{p}
\,,\left(\frac{p_{b}}{p}\right)^{2-\frac{2\,\varepsilon}{3}}\right)
\end{eqnarray}
In (\ref{rg:energy}) the wave number
\begin{eqnarray}
p_{b}\propto\left(\frac{\varepsilon}{\nu^{3}\,\tau^{3}}\right)^{\frac{1}{6-2\,\varepsilon}}
\end{eqnarray}
signals whether at small scales dissipation is mainly
due to friction ($p \ll p_{b}$)  or to molecular viscosity ( $p \gg p_{b}$ ).  
The function $R$ has a regular expansion in $\varepsilon$ for fixed $p_{b}$. 
The infra-red scaling of (\ref{rg:energy}) is then determined by the 
behavior of $R$ in the limit $p\downarrow 0$. This limit is inquired
within the renormalization group formalism by the so-called operator product 
expansion. The outcome \cite{Ho98} is that the energy spectrum admits 
the same infra-red asymptotics as in dimension higher than two
\begin{eqnarray}
\label{rg:energyasy}
E_{[\boldsymbol{v}]}(p)\sim p^{1-\frac{4\,\varepsilon}{3}}
\end{eqnarray} 
It is worth emphasizing that the above results were derived in \cite{Ho98} using 
the vorticity representation of the velocity field which holds only in $2\,d$. 
As a further check, in \cite{HoKaKo02,HnHoHoSe99} the same results were recovered 
by analytic continuation of (\ref{model:ns}) in the limit $d\,\downarrow\,2$. 
In \cite{HnHoHoSe99} the analysis extends to the scaling properties of the 
concentration field and gives
\begin{eqnarray}
\label{rg:enstrophysp}
\hspace{-1.0cm}
E_{[\theta]}(p):=\int\frac{d^{2}q}{(2\,\pi)^{2}}\,\delta(p-q)\int d^{2}x
\,e^{\imath\boldsymbol{p\cdot x}}\prec\theta(\boldsymbol{x},t)\theta(0,t)\succ 
\sim p^{1+\frac{2\,\eps}{3}-2\,h}
\end{eqnarray}
in agreement with the dimensional prediction (\ref{rg:inertial}). 
In summary, according to the renormalization group analysis of 
\cite{Ho98,HoKaKo02,HnHoHoSe99}, in $2\,d$ as in $3\,d$
the scaling properties of (\ref{model:ns}), (\ref{model:ps}) differ 
for
$\varepsilon$ tending to zero from those of fully developed turbulence.
In particular, inverse cascade-like scaling is attained only for
$\varepsilon$ equal two and direct cascade-like at $\varepsilon$
equal three. In the following section we will argue that these
results are in contradiction  with those that a phenomenological theory
 \emph{\`a la} Kraichnan would suggest.

\section{Phenomenology \emph{\`a la} Kraichnan}
\label{sec:Kraichnan}

Our starting point are the K\'arm\'an-Howarth-Monin equation \cite{Frisch} 
\begin{eqnarray}
\label{Kraichnan:KHM}
\hspace{-1.5cm}\lefteqn{\frac{1}{2}\de_{\mu}\prec\flu{v}^{\mu}(\boldsymbol{x},t)\, 
||\flu{\boldsymbol{v}}(\boldsymbol{x},t)||^{2}\succ=}
\nonumber\\
\hspace{-0.8cm}\left(\de_{t}+\frac{2}{\tau}\right)\prec v^{\alpha}(\boldsymbol{x},t)
v_{\alpha}(\boldsymbol{0},t)\succ
+2\,\nu\,\prec \de_{\mu}v^{\alpha}(\boldsymbol{x},t) 
\de^{\mu} v_{\alpha}(\boldsymbol{0},t)\succ
-F^{\alpha}_{\hspace{0.2cm}\alpha}(\boldsymbol{x})
\end{eqnarray}
and the analogous expression for the scalar field
\begin{eqnarray}
\label{Kraichnan:KHMc}
\hspace{-1.5cm}\lefteqn{\frac{1}{2}\de_{\mu}\prec\flu{v}^{\mu}(\boldsymbol{x},t) 
\sbr{\flu{\theta}(\boldsymbol{x},t)}^{2}\succ=}
\nonumber\\
\hspace{-0.5cm}\de_{t}\prec[\flu{\theta}(\boldsymbol{x},t)]^{2}\succ
+2\,\kappa\,\prec\de_{\mu}\theta(\boldsymbol{x},t) 
\de^{\mu} \theta(\boldsymbol{0},t)\succ-G(\boldsymbol{x})
\end{eqnarray}
In (\ref{Kraichnan:KHM}) and (\ref{Kraichnan:KHMc}) the notation is
\begin{eqnarray}
&&\flu{v}^{\mu}(\boldsymbol{x},t) :=v^{\mu}(\boldsymbol{x},t) -v^{\mu}(\boldsymbol{0},t)
\nn
&&\flu{\theta}(\boldsymbol{x},t):=\flu{\theta}(\boldsymbol{x},t) -\flu{\theta}(\boldsymbol{0},t)
\end{eqnarray}
The scaling predictions of Kraichnan's theory stem from the 
asymptotic solution of (\ref{Kraichnan:KHM}) under the following 
three assumptions \cite{Be00}:
\begin{enumerate}
\item[$(i)$] velocity correlations are smooth at finite viscosity and 
exist in the inviscid limit even at coinciding points,
\item[$(ii)$] even in the absence of large scale friction (i.e. $\tau=\infty$) 
Galilean invariant functions, and in particular structure functions, 
reach a steady state,
\item[$(iii)$] no dissipative anomalies occur for the energy cascade.
\end{enumerate}
The assumptions $(i)$, $(ii)$ imply that the two point correlation in the 
absence of Ekman friction does not reach a steady state
\begin{eqnarray}
\prec v^{\alpha}(\boldsymbol{x},t)\, v_{\alpha}(\boldsymbol{0},t)\succ
=\lambda\,t-\frac{1}{2}\prec||\delta\boldsymbol{v}(\boldsymbol{x},t)||^{2}\succ
+\dots
\hspace{1.0cm} (\tau=\infty)
\label{Kraichnan:nonsteady}
\end{eqnarray}
the constant $\lambda$ being the asymptotic growth-rate. 
By $(iii)$ energy dissipation in (\ref{Kraichnan:KHM}) satisfies
\begin{eqnarray}
\label{Kraichnan:noda}
\left\{\lim_{\nu\downarrow 0}\lim_{x\downarrow 0}-\lim_{x\downarrow 0}\lim_{\nu\downarrow 0}\right\}
\nu\, \prec\de_{\mu} v^{\alpha}(\boldsymbol{x},t)\de^{\mu} v_{\alpha}(\boldsymbol{0},t)\succ=0
\end{eqnarray}
This latter hypothesis is distinctive of two dimensional turbulence: 
in three and higher dimensions the limits are not expected to commute 
for fully developed turbulence. If the bulk of the energy injection $I_{\mathcal{E}}$
occurs around a wave-number $p_{f}$ and viscosity and friction are such that the 
adimensional parameter
\begin{eqnarray}
\label{Kraichnan:rey}
\mathcal{R}=\frac{I_{\mathcal{E}}\,\tau^{2}}{\nu}\,\gg\,1
\end{eqnarray}
plays the role of a large Reynolds number then the three
hypotheses yield an inverse cascade for 
wave-numbers $\boldsymbol{p}$ in the range $p_{\tau}\,\ll\,p\,\ll\,p_{f}$ 
with $p_{\tau}=(I_{\mathcal{E}}\,\tau^{3})^{-1/2}$ and a direct cascade for 
$p_{f}\,\ll\,p\,\ll\,\bar{p}_{\tau}$ 
with $\bar{p}_{\tau}=(\nu\,\tau)^{-1/2}=p_{\tau}/\mathcal{R}^{1/2}$ \cite{Be99a,Be99b}. 
Note that the Kolmogorov scale $p_{K}=:(I_{\mathcal{E}}/\nu^{3})^{1/4}=p_{\tau}/\mathcal{R}^{3/4}$ 
is always smaller than the dissipation scale set by the Ekman friction.
We show below how the same arguments can be adapted to a power law forcing. 

\subsection{Inverse cascade:  $\varepsilon<2$}
\label{sec:inverse}
By (\ref{model:Einjection}) and (\ref{model:Zinjection}) the energy and 
the enstrophy input are in this case dominated by the ultra-violet 
cut-off $M_{f}$. Neglecting $m_{f}$, the trace of the forcing correlation 
admits the asymptotic expansion (see \ref{ap:forcing} for details)
\begin{eqnarray}
\label{model:forcingasy}
\hspace{-1.5cm}
F_{\hspace{0.2cm}\alpha}^{\alpha}(\boldsymbol{x},0,M_{f})=\left\{
\begin{array}{ll}
M_{f}^{4-2\,\varepsilon}\left\{I_{\mathcal{E}}(0,1)
-\frac{I_{\mathcal{Z}}(0,1)\,(M_{f}\,x)^{2}}{2}+\dots\right\}\,
\hspace{0.2cm}&\hspace{0.2cm} M_{f} x \,\ll\,1
\nonumber\\[0.5cm]
\frac{4^{1-\varepsilon}\,\Gamma\left(2-\varepsilon\right)}{\pi\,
\Gamma(\varepsilon-1)}\frac{F_{o}}{x^{4-2\,\varepsilon}}\,
\hspace{0.2cm}&\hspace{0.2cm} M_{f} x \,\gg\,1
\end{array}
\right.
\end{eqnarray}
holding for $0\,<\,\varepsilon\,<2$. Comparison with the renormalization group results is 
possible in the infra-red region, $M_{f}x\,\gg\,1$. In order to extricate the corresponding 
asymptotics of the third order structure function, it is convenient to consider 
first the quasi stationary case for $\tau$ tending to infinity. 
By hypotheses $(i)$ and $(iii)$ \cite{Be99a}, %the inviscid limit at 
%coinciding points of (\ref{Kraichnan:KHM}) equals 
the asymptotic growth-rate of (\ref{Kraichnan:nonsteady}) in the inviscid limit 
is equal to to the energy injection
\begin{eqnarray}
\label{Kraichnan:injection}
\lambda=%I_{\mathcal{E}}\propto F_{0}\,M_{f}^{4-2\,\varepsilon}
M_{f}^{4-2\,\varepsilon}\,I_{\mathcal{E}}(0,1):=M_{f}^{4-2\,\varepsilon}\,F_{o}^{\star}
\end{eqnarray}
By (\ref{Kraichnan:KHM}) the growth-rate sustains the structure function 
at scales $M_{f} x \gg 1$
\begin{eqnarray}
\label{Kraichnan:inverse}
\hspace{-1.5cm}
\prec\delta v^{\mu}(\boldsymbol{x},t)||\delta v||^{2}(\boldsymbol{x},t)\succ
=F_{o}^{\star}\,M_{f}^{4-2\, \varepsilon}\,x^{\mu}
%I_{\mathcal{E}}(0,M_{f})
\left\{1-\frac{4^{1-\varepsilon}\,
\Gamma\left(2-\varepsilon\right)\,F_{o}}{\pi\,
\Gamma(\varepsilon)\,F_{o}^{\star}\,(M_{f}x)^{4-2\,\varepsilon}}+\dots\right\}
\end{eqnarray} 
In the presence of the Ekman friction ($\tau\,<\,\infty$) (\ref{Kraichnan:KHM}) reaches 
at steady state. In such a case the energy injection is balanced 
by the velocity correlation which far from the infra-red cut-off
 $p_{\tau}=(F_{o}\,M_{f}^{4-2\,\varepsilon}\tau^{3})^{-1/2}$ is expected to take the form
\begin{eqnarray}
\label{Kraichnan:twopoints}
\hspace{-0.8cm}
\prec \boldsymbol{v}^{\alpha}(\boldsymbol{x},t)\boldsymbol{v}_{\alpha}(\boldsymbol{0},t)\succ
=
\frac{\tau\,F_{0}^{\star}\,M_{f}^{4-2\,\varepsilon}}{2}\left\{1
-c_{1}\,\left(p_{\tau}\,x\right)^{\zeta_{2}}+\dots\right\}
\end{eqnarray}
with $c_{1}$ a pure number and $\zeta_{2}$ to be determined by a
self-consistence condition.
The asymptotics of the structure function acquires a correction
\begin{eqnarray}
\label{Kraichnan:inversefriction}
\hspace{-1.0cm}
\lefteqn{
\prec\delta v^{\mu}(\boldsymbol{x},t)||\delta v||^{2}(\boldsymbol{x},t)\succ=}
\nonumber\\
&&
F_{o}^{\star}\,M_{f}^{4-2\,\varepsilon}\,x^{\mu}
\left\{1-\frac{2\,c_{1}\,(p_{\tau}\,x)^{\zeta_{2}}}{(2+\zeta_{2})}
-\frac{4^{1-\varepsilon}\,\Gamma\left(2-\varepsilon\right)\,F_{o}}{\pi\,
\Gamma(\varepsilon)\,F_{o}^{\star}\,(M_{f}x)^{4-2\,\varepsilon}}+\dots\right\}
\end{eqnarray}
%with $c_{1}$ a pure number and $p_{\tau}=(F_{o}\,M_{f}^{4-2\,\varepsilon}\tau^{3})^{-1/2}$. 
Some remarks are in order. The constant flux solution dominates for
\begin{eqnarray}
\label{Kraichnan:inverserange}
\frac{1}{M_{f}}\,\ll\,x\,\ll\,(F_{o}M^{4-2\,\varepsilon}\tau^{3})^{1/2}
\end{eqnarray}
In this range the renormalization group prediction clearly appears 
as a \emph{sub-leading} correction. Similarly, the two-point correlation
adds a further sub-leading term associated to the exponent $\zeta_{2}$. 
Dimensional considerations yield for $\zeta_{2}$ the value $2/3$ whence
a $-5/3$ exponent follows for the energy spectrum. The sign of the constant
flux term stemming from (\ref{Kraichnan:inverse}), (\ref{Kraichnan:inversefriction}) 
is positive so describing energy transfer to larger scales. The conclusion
is of an inverse cascade taking place above the forcing ultra-violet
cut-off. In such a case, the study of the statistics of the passive scalar 
should recover the results of \cite{BiCeLaSbTo04}. A priori inspection of
(\ref{Kraichnan:KHMc}) allows one to distinguish at least two sub-cases.

\subsubsection{Scalar field in the inverse cascade and small scale forcing $h\,<\,2$.}
\label{sec:cinversesmall}

For $h\,<\,2$ the injection of scalar fluctuations is concentrated in the ultra-violet 
(thermal stirring). No dissipative anomaly is expected.  
The K\'arm\'an-Howarth-Monin equation (\ref{Kraichnan:KHMc}) 
yields 
\begin{eqnarray}
\label{Kraichnan:KHMclesstwosol}
\prec\flu{v}^{\mu}(\boldsymbol{x},t) 
\left[\flu{\theta}(\boldsymbol{x},t)\right]^{2}\succ \simeq
-\frac{4^{1-h}\,\Gamma\left(2-h\right)\,G_{o}\,x^{\mu}}{\pi\,
\Gamma(h)\,x^{4-2\,h}}
\end{eqnarray}
in the scaling range
\begin{eqnarray}
\tilde{m}=\left\{m_{f}^{-1},m_{g}^{-1}\right\}\,\gg\,x\,\gg\,
\mathrm{min}\left\{M_{f}^{-1},M_{g}^{-1}\right\}:=\tilde{M}
\end{eqnarray}
Inferences about the correlation functions of the scalar field can
be barely drawn from crude dimensional considerations. Accordingly, one has
\begin{eqnarray}
\label{Kraichnan:cspectrum}
E_{[\theta]}(p)\sim \frac{\prec\left[
\delta\theta\left(\frac{\boldsymbol{p}}{p^{2}},t\right)\right]^{2}\succ}{p^{1-\frac{1}{3}}}
\sim p^{\frac{7-6\,h}{3}}
\end{eqnarray}
which differs from the one stemming from the scaling dimension
$d_{\theta}$ given in (\ref{rg:inertial}) and supported by the 
renormalization group calculations of \cite{HnHoHoSe99}.  
In particular, whilst (\ref{rg:inertial}) recovers equipartition 
scaling only for $(\varepsilon,h)=(0,0)$, (\ref{Kraichnan:cspectrum})
yields a scaling linear in wave-number space at $h=2/3$. 
This may indicate the breakdown for $h\,<\,2/3$ of 
(\ref{Kraichnan:cspectrum}) and the onset of an equipartition-type 
scaling for the spectrum of the scalar field, see \ref{ap:KM}
for quantitative modeling of the phenomenon.

\subsubsection{Scalar field in the inverse cascade and large scale forcing $h\,>\,2$.}
\label{sec:cinverselarge}

In this regime, the injection of the scalar field is dominated by the
infra-red cut-off. Physically the situation may be assimilated
to turbulent stirring.
The K\'arm\'an-Howarth-Monin equation for $m_{g}\,x\,\ll\,1$ 
reduces to
\begin{eqnarray}
\label{cinverse:largeeq}
\hspace{-1.0cm}
\lefteqn{\frac{1}{2}\de_{\mu}\prec v^{\mu}(\boldsymbol{x},t)
\left[\delta \theta\left(\boldsymbol{x},t\right)\right]^{2}\succ \simeq}
\nonumber\\&&
-\,G_{o}\,m_{g}^{4-2\,h}\left\{\bar{\gamma}_{h-2}+(m_{g}x)^{2\,h-4}\bar{\gamma}_{0}
+(m_{g}x)^{2}\bar{\gamma}_{h-1}+\dots\right\}
\end{eqnarray}
The coefficients of the forcing expansion are specified by the 
formulae of \ref{ap:forcing} by identifying $\gamma$ with the 
function $\phi$ thereby defined. Equation (\ref{cinverse:largeeq}) 
points at a direct cascade of the scalar field with sub-leading 
corrections due to the power-law forcing. The dimensional prediction 
for the spectrum of the scalar field is Obukhov-Corssin's \cite{Frisch}
\begin{eqnarray}
\label{Kraichnan:OC}
E_{[\theta]}(p)\sim p^{-5/3}
\end{eqnarray}
The numerical experiments of \cite{BiCeLaSbTo04} support  Obukhov-Corssin's
scaling in this regime.

\subsection{Local balance: $2\,<\varepsilon\,<\,3$}
\label{sec:localbalance}

For $2\,<\,\eps\,<\,3$ the injection of the scalar field is dominated by the infra-red
cut-off $m_{f}$. The quasi-steady state solution for vanishing Ekman
friction yields
\begin{eqnarray}
\label{Kraichnan:balanceinjection}
\lambda=m^{4-2\,\varepsilon}\,I_{\mathcal{E}}(1,0):=F_{o}\,m_{f}^{4-2\,\varepsilon}\,
\phi_{\varepsilon-2}
\end{eqnarray}
Correspondingly, the scaling range is set by the condition $m_{f}\,x\ll\,1$
alone and (\ref{Kraichnan:KHM}) becomes
\begin{eqnarray}
\label{localbalance:eq}
\hspace{-1.0cm}
\lefteqn{\frac{1}{2}\de_{\mu}\prec\delta v^{\mu}(\boldsymbol{x},t)\, 
||\flu{\boldsymbol{v}}(\boldsymbol{x},t)||^{2}\succ =}
\nonumber\\
&&
\lambda-F_{o}\,m_{f}^{4-2\,\varepsilon}\,\left\{\phi_{\varepsilon-2}+
(m_{f}\,x)^{2\,\varepsilon-4}\phi_{0}+(m_{f}\,x)^{2}\phi_{\varepsilon-1}+\dots\right\}
\end{eqnarray}
As the first two terms on the right hand side of (\ref{localbalance:eq})
cancel out, the third order structure function admits the asymptotic
expression
\begin{eqnarray}
\label{localbalance:sf}
\hspace{-1.0cm}
\prec\delta v^{\mu}(\boldsymbol{x},t)\,||\flu{\boldsymbol{v}}(\boldsymbol{x},t)||^{2}\succ
\simeq 
-F_{o}\,x^{\mu}\,x^{2\,\eps-4}\cbr{\frac{\phi_{0}}{\varepsilon-1}
+\frac{(m_{f}\,x)^{6-2\,\eps}\,\phi_{\varepsilon-1}}{2}+\dots}
\end{eqnarray}
with $\phi_{0}\,<\,0$ (see \ref{ap:forcing}). The steady state solution in the
presence of Ekman friction can be discussed as in subsection~\ref{sec:inverse}
and introduces in such a case only sub-leading terms. The cancellation 
in (\ref{localbalance:eq}) was argued in \cite{Be99a,Be99b,Be00} to 
underlie the direct cascade for standard turbulent forcing. The difference
here is that power-law forcing dominates the scaling. In this range, the
solution of (\ref{Kraichnan:KHM}) validates the dimensional prediction 
(\ref{rg:inertial}). The agreement extends to the scalar field
with two provisos
\begin{enumerate}
\label{localbalance}
\item $h\,<\,2$: the threshold for equipartition scaling is 
$h^{\star}=\varepsilon/3>2/3$:
\item $h>2$: the forcing becomes dominated by the infra-red cut-off.
Correspondingly, a "\emph{freezing}" of the scaling dimensions at the
value for $h=2$ may be expected \cite{FoFr83,AdAnVa}.
\end{enumerate} 
In summary, the expected spectra are
\begin{eqnarray}
\label{localbalance:spectra}
E_{[\boldsymbol{v}]}(p)\sim p^{1-\frac{4\,\varepsilon}{3}}
\hspace{1.0cm}\&\hspace{1.0cm}
E_{[\theta]}(p)\sim\left\{
\begin{array}{ll}
p\,, \hspace{0.3cm} & \hspace{0.3cm} 0\,<\,h\,<\,\frac{\varepsilon}{3}
\\
p^{1+\frac{2\,\varepsilon}{3}-2\,h}\,, 
\hspace{0.3cm} & \hspace{0.3cm} \frac{\varepsilon}{3}\,<\,h\,<\,2
\\
p^{-3+\frac{2\,\varepsilon}{3}}\,, \hspace{0.3cm} & \hspace{0.3cm}  h\,>\,2
\end{array}
\right.
\end{eqnarray}
It should be noted that in this regime for $\varepsilon/3\,<\,h\,<\,2$ the predictions of 
the renormalization group and of the phenomenological theory coincide. 

\subsection{Direct cascade : $\eps\,>\,3$}
\label{sec:direct}

For $\varepsilon\,>\,3$ both energy (\ref{model:Einjection}) 
and enstrophy (\ref{model:Zinjection}) injection are dominated
by the infra-red cut-off. The analysis of the K\'arm\'an-Howarth-Monin
equation under $(i),(ii),(iii)$ in the quasi-steady state follows 
the same lines as in previous subsection~\ref{sec:localbalance}. 
The second term in the square brackets of (\ref{localbalance:eq})
now dominates scaling
\begin{eqnarray}
\label{direct:sf}
\hspace{-1.0cm}
\prec\delta v^{\mu}(\boldsymbol{x},t)\,||\flu{\boldsymbol{v}}(\boldsymbol{x},t)||^{2}\succ
\simeq 
-F_{o}\,x^{\mu}\,x^{2}\,m_{f}^{6-2\varepsilon}\cbr{
\frac{\phi_{\varepsilon-1}}{2}+\frac{(m_{f} x)^{2\,\varepsilon-6}\phi_{0}}{\varepsilon-1}+\dots}
\end{eqnarray}
However, (\ref{direct:sf}) is just the simplest of the possible 
scenarios. A detailed analysis of the direct cascade stirred by
turbulent forcing \cite{Be99b} indicates that scaling in the steady 
state brought about by an Ekman friction may well be characterized 
by \emph{non-universal} exponents depending upon the value of $\tau$. 
Subsequent numerical investigations \cite{NaOtAnGu00,BoCeMuVe02,TsOtAnGu05}
validate non-universal exponents in the steady state. A further
inference drawn in \cite{Be99b} is the presence of logarithmic
corrections to the quasi-steady state structure function. For the
scopes of the present work it is sufficient to observe that the
phenomenological theory predicts for $\varepsilon\,>\,3$ the 
"freezing" of the scaling dimension $d_{v}$ to a value close
to the "naive" direct cascade ($d_{v}=d_{x}$). The implication for the
energy spectra by dimensional arguments is
\begin{eqnarray}
\label{direct:spectra}
E_{[\boldsymbol{v}]}(p)\sim p^{-3+\dots}
\hspace{1.0cm}\& \hspace{1.0cm}
E_{[\theta]}(p)\sim 
\left\{
\begin{array}{ll}
p\,, \hspace{0.3cm} & \hspace{0.3cm} 0\,<\,h\,< \,1
\\
p^{3-2\,h+\dots}\,, \hspace{0.3cm} & \hspace{0.3cm} 1\,<\,h\,< \,2 
\\
p^{-1+\dots}\,, \hspace{0.3cm} & \hspace{0.3cm} h\,\geq \,2 
\end{array}
\right.
\end{eqnarray}
In (\ref{direct:spectra}) the "$\dots$" denote non-universal and/or 
intermittent correction the presence whereof is suggested by 
numerical investigations of the direct cascade stirred by turbulent forcing
\cite{NaOtAnGu00,BoCeMuVe02,TsOtAnGu05}.

\subsection*{Observation:}
In the above discussion we omitted  to discuss the properties of the
flow of (\ref{model:ns}), (\ref{model:ps}) at scales $M_{f}\, x\,\ll\,1$.
The reason for doing so, in the spirit of renormalization group and
consistently with the numerical experiments of the ensuing section,
is that the dynamics is strongly  suppressed by dissipation
effects for scales below $M_{f}$.

\section{Numerical experiments}
\label{sec:numerics}

In order to compare 
the renormalization group predictions of section~\ref{sec:rg}
with those of the phenomenological theory \emph{\`a la} Kraichnan 
expounded in section~\ref{sec:Kraichnan}, we performed numerical simulations
of the Navier--Stokes equation for the vorticity field (\ref{model:vorticity})
and the advection-diffusion equation for the scalar field (\ref{model:ns})
with a fully-dealiased pseudo-spectral method \cite{CaHuQuZa} in a doubly 
periodic square domain of size $L=2\pi$ at resolution $N^2 = 1024^2$. 
Dealiasing cutoff is set to $k_t= N/3$. 
Time evolution was computed by means of a second-order Runge--Kutta scheme, 
with implicit handling of the linear friction and viscous terms. 
As customary (see e.g. \cite{NaOtAnGu00,TsOtAnGu05}) we added to (\ref{model:ns}) 
an hyperviscous damping $(-1)^{p-1}\nu_{p-1}\de^{2\,p}\,v$.This is equivalent to a 
Pauli-Villars regularization the use whereof is well justified in 
renormalization group calculations (see e.g. \cite{Zinn,Ho98}). 
The integration time has been carried out for twenty large eddy turn-over 
times after the velocity fields have reached the stationary state.
The stochastic forcing is implemented in Fourier space by means of Gaussian, 
white-in-time noise as in \cite{BoCeMuVe02} but with variance determined 
according to (\ref{model:F}).

%%%%%%%%%%%%%%%%%%%%%%%%%%%%%%%%%%%%%%%%%%%%%%%%%%%%%%%%%%%%%%%%%%%%%%%%%%%%%%%%%%%%%%%%%%%%
 \begin{figure}[top]
  \centering
  \includegraphics[width=10cm]{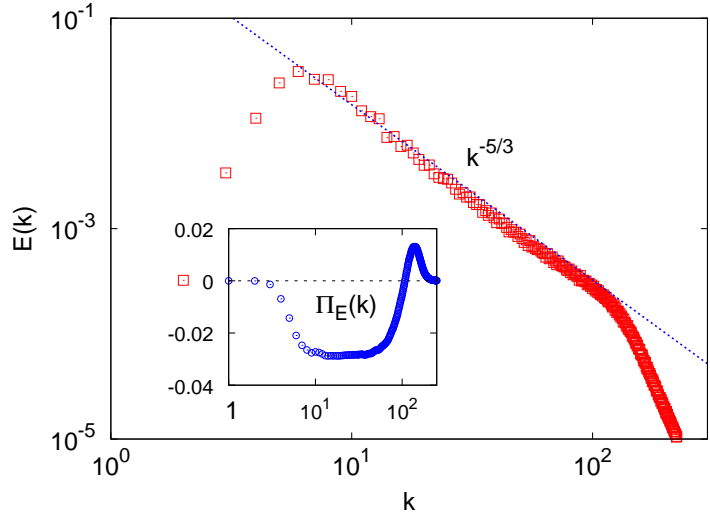}
  \caption{Kinetic energy spectrum for $\varepsilon=0$. 
    Inset: energy flux $\Pi_E$. 
    Parameters values are: 
    $m_f=1,M_f=341$, $\nu_{3}=10^{-16}$, $\tau^{-1}_2=10^2$, $F_0=4\cdot10^{-10}$}
  \label{fig:sp_e0}
\end{figure}
%%%%%%%%%%%%%%%%%%%%%%%%%%%%%%%%%%%%%%%%%%%%%%%%%%%%%%%%%%%%%%%%%%%%%%%%%%%%%%%%%%%%%%%%%%%%
 \begin{figure}[top]
  \centering
  \includegraphics[width=10.0cm]{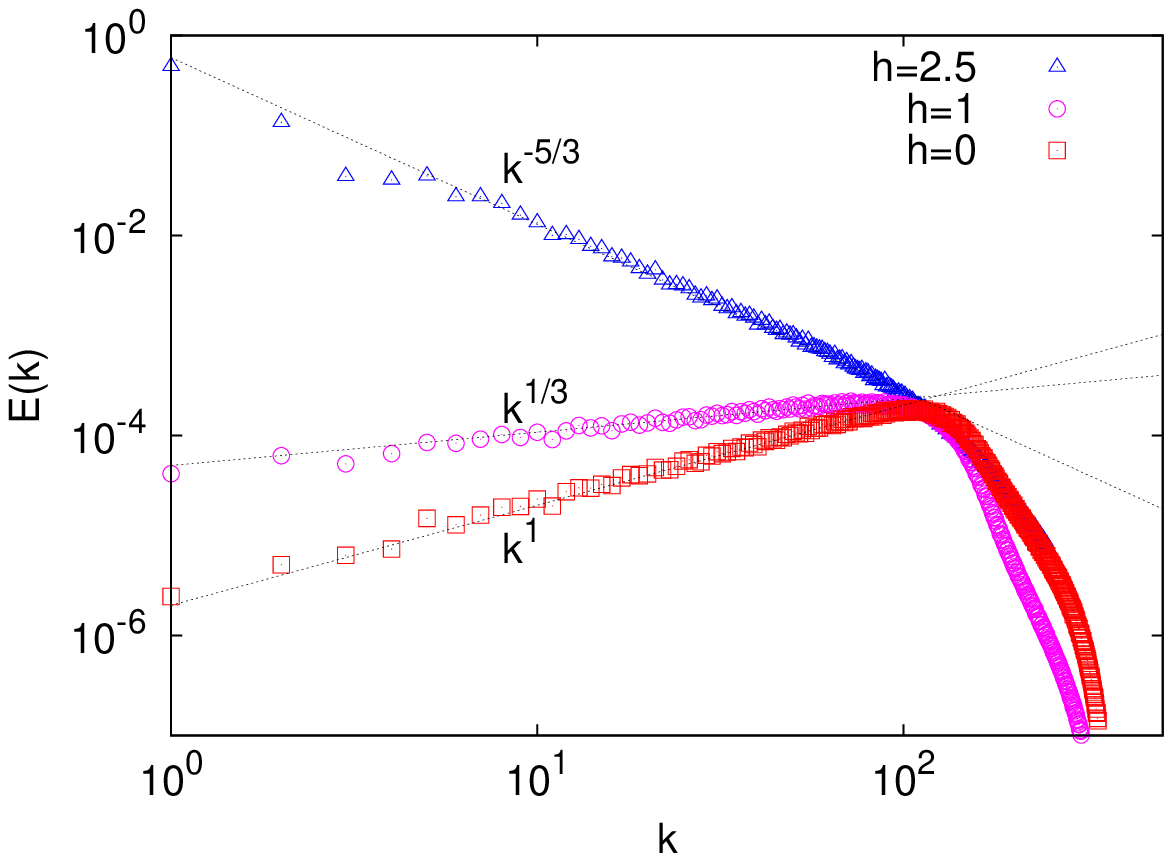}
  \caption{Scalar spectra for $\varepsilon=0$ and various values of $h$. 
    Parameters values are:
    $m_g=1,M_g=341$, $\kappa_{3}=10^{-16}$, $G_0=4\cdot10^{-10}$ for $h=0$,
    $m_g=1,M_g=341$, $\kappa_{3}=10^{-16}$, $G_0=1.6\cdot10^{-5}$ for $h=1$,
    $m_g=1,M_g=341$, $\kappa_{0}=5\cdot10^{-4}$, $G_0=2.5\cdot10^{-1}$ for $h=2.5$.
    Parameters for the velocity field as in Fig.~\ref{fig:sp_e0}
   }
  \label{fig:spt_e0}
\end{figure}
%%%%%%%%%%%%%%%%%%%%%%%%%%%%%%%%%%%%%%%%%%%%%%%%%%%%%%%%%%%%%%%%%%%%%%%%%%%%%%%%%%%%%%%%%%%%

%%%%%%%%%%%%%%%%%%%%%%%%%%%%%%%%%%%%%%%%%%%%%%%%%%%%%%%%%%%%%%%%%%%%%%%%%%%%%%%%%%%%%%%%%%%%
 \begin{figure}[top]
  \centering
  \includegraphics[width=10cm]{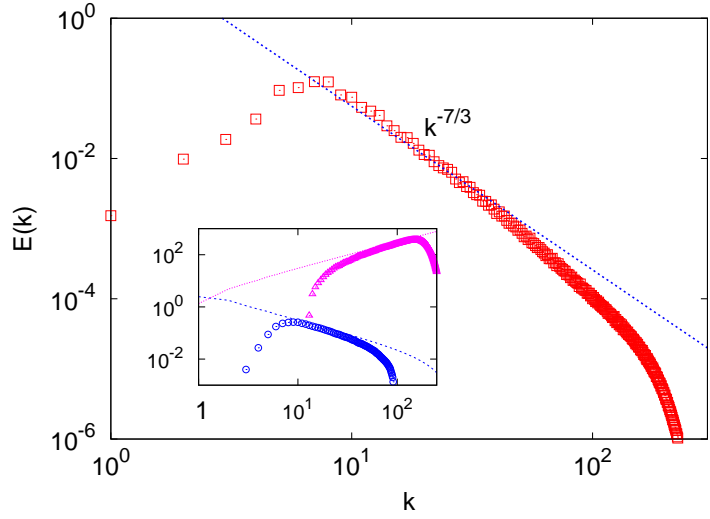}
  \caption{Kinetic energy spectrum $E(k)$ for $\varepsilon=2.5$. 
    Inset: energy flux $\Pi_E$ (circles) multiplied by minus the unity, 
    and enstrophy flux $\Pi_Z$ (triangles). 
    The lines represents the injection spectra $I_E$ (dashed line) and $I_Z$ (dotted line). 
    Parameters values are: 
    $m_f=1,M_f=341$, $\nu_{3}=10^{-17}$, $\tau^{-1}_2=10^3$, $F_0=1$}
  \label{fig:sp_e2.5}
\end{figure}
%%%%%%%%%%%%%%%%%%%%%%%%%%%%%%%%%%%%%%%%%%%%%%%%%%%%%%%%%%%%%%%%%%%%%%%%%%%%%%%%%%%%%%%%%%%%
 \begin{figure}[top]
  \centering
  \includegraphics[width=10.0cm]{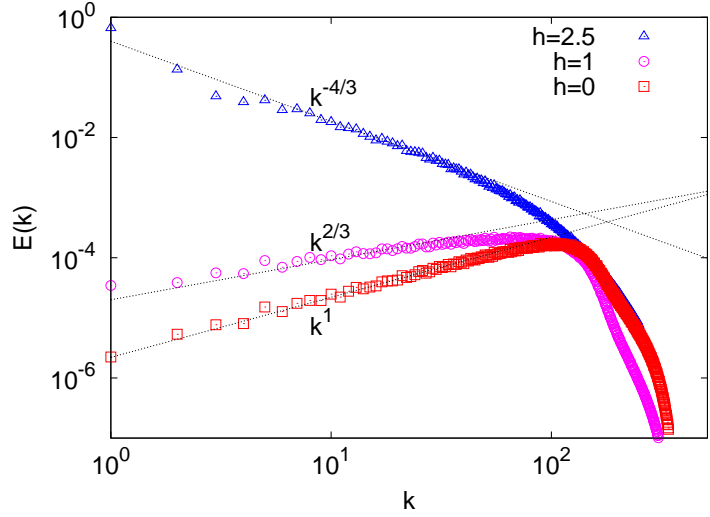}
  \caption{Scalar spectra for $\varepsilon=0$ and various values of $h$. 
    Parameters values are:
    $m_g=1,M_g=341$, $\kappa_{3}=10^{-16}$, $G_0=4\cdot10^{-10}$ for $h=0$,
    $m_g=1,M_g=341$, $\kappa_{3}=10^{-16}$, $G_0=1.6\cdot10^{-5}$ for $h=1$,
    $m_g=1,M_g=341$, $\kappa_{0}=5\cdot10^{-4}$, $G_0=2.5\cdot10^{-1}$ for $h=2.5$.
    Parameters for the velocity field as in Fig.~\ref{fig:sp_e2.5}
   }
  \label{fig:spt_e2.5}
\end{figure}
%%%%%%%%%%%%%%%%%%%%%%%%%%%%%%%%%%%%%%%%%%%%%%%%%%%%%%%%%%%%%%%%%%%%%%%%%%%%%%%%%%%%%%%%%%%%

In Fig.~\ref{fig:sp_e0} we show the energy spectrum $E_{[\boldsymbol{v}]}$ for
$\varepsilon$ equal zero. The numerical spectrum exhibits good agreement
with the phenomenological theory of section~\ref{sec:Kraichnan} with a scaling
exponent $d_{\mathcal{E}}=-5/3\,d_{p}$ within numerical accuracy.

%{\bf Non capisco bene cosa significa $d_p$... Anche $d_{\mathcal{E}}$ e un po brutto come simbolo perche ${\mathcal{E}}$ si confonde con $\varepsilon$}

Such an exponent 
is very far from the equipartition-like scaling $d_{\mathcal{E}}=d_{p}$ which is the 
starting point for the renormalization group analysis. 
The energy flux validates the interpretation of 
the $-5/3$ spectrum as brought about by an inverse cascade. 
The energy flux 
\begin{eqnarray}
\label{numerics:flux}
\Pi_{E}(p,m_{f})=\int_{m_{f}}^{p}\frac{d^{2}q}{(2\,\pi)^{2}}\,
\Re\,\int d^{2}x\,e^{\imath \boldsymbol{q\cdot x}}
\prec \check{v}^{\alpha}(-\boldsymbol{q},t)
(v^{\beta}\de_{\beta} v_{\alpha})(\boldsymbol{x},t)\succ
\end{eqnarray}
where $\check{v}^{\alpha}$ denotes the Fourier transform of $v^{\alpha}$, 
is \emph{negative} and constant in the scaling range (see inset of
Fig.~\ref{fig:sp_e0}), so signaling the presence of an inverse cascade.

Breakdown of 
the marginality assumption at $\varepsilon$ equal zero is confirmed also 
by the concentration spectra $E_{[\theta]}$ shown in Fig.~\ref{fig:spt_e0}: 
at $h=1$ a spectrum $E_{[\theta]}\sim p^{1/3}$ well fits the numerical results,
which definitely rule out the $E_{[\theta]}\sim p^{-1}$ prediction of the 
renormalization group theory (\ref{rg:enstrophysp}). 
Furthermore and in agreement with sub-section~\ref{sec:inverse} the 
spectrum of the scalar field undergoes a transition at $h=2$. 
Above that value the scaling exponent ¨freeze¨ to the value $-5/3$
corresponding to a direct cascade, in agreement with the results 
of \cite{BiCeLaSbTo04}. Fig.~\ref{fig:spt_e0} illustrates the phenomenon 
for $h$ equal $2.5$. 
It should be noted that for $h$ equal zero an 
equipartition-type scaling (i.e. linear in wave-number) is observed for the 
spectrum of the scalar field. It is worth stressing that this is not sufficient
to infer equipartition of the full statistics of the scalar field. 
It is known in analytically tractable cases of turbulent advection 
of scalar fields\cite{FaFo05,CeSe05,CeSe06,MaMG09} that equipartition-type scaling
of the two-point correlation may well co-exist with highly intermittent
statistics even in the decay range of a scalar field.

In agreement with the phenomenological theory, for $2\,<\,\varepsilon\,<\,3$
no cascade is observed. 
Fig.~\ref{fig:sp_e2.5} illustrates the situation at $\varepsilon=2.5$. 
The energy injection spectrum, defined as 
$
I_{E}(k)=\int_{k}^{\infty}dp\,p^{3-2\,\varepsilon}\chi_{f}\left(\frac{m_{f}^{2}}{p^{2}}\,,
\frac{p^{2}}{M_{f}^{2}}\right) 
$ 
is dominated by $IR$ contributions, 
while the enstrophy injection spectrum
$
I_{Z}(k)=\int_{0}^{k}dp\,p^{5-2\,\varepsilon}\chi_{f}\left(\frac{m_{f}^{2}}{p^{2}}\,,
\frac{p^{2}}{M_{f}^{2}}\right) 
$  
is still ultraviolet divergent (see inset of Fig.~\ref{fig:sp_e2.5}). 
In this situation the steady state is characterized by a scale-by-scale balance 
between the fluxes and the injection spectra. The resulting 
energy spectrum scales as 
$E_{[\boldsymbol{v}]}(p)\sim p^{1-\frac{4\,\varepsilon}{3}}$ (see Fig.~\ref{fig:sp_e2.5}).
The spectra for the passive scalar shown in Fig.~\ref{fig:spt_e2.5}
are in agreement with the prediction (\ref{localbalance:spectra}). 

%%%%%%%%%%%%%%%%%%%%%%%%%%%%%%%%%%%%%%%%%%%%%%%%%%%%%%%%%%%%%%%%%%%%%%%%%%%%%%%%%%%%%%%%%%%%
 \begin{figure}[top]
  \centering
  \includegraphics[width=10cm]{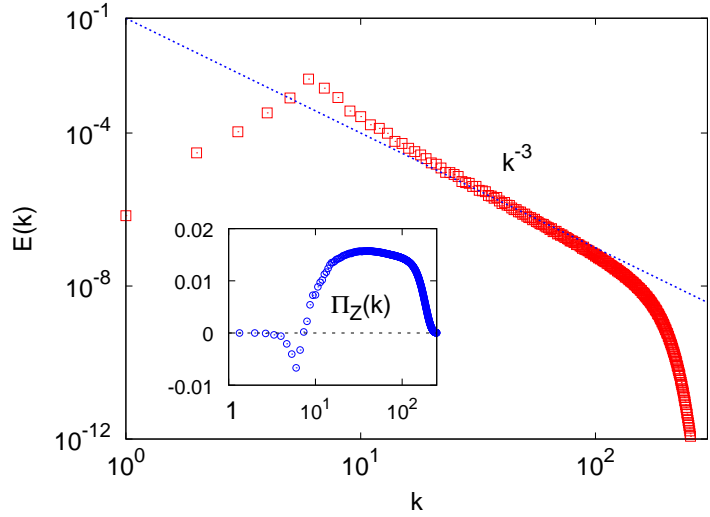}
  \caption{Kinetic energy spectrum for $\varepsilon=4$. 
    Inset: enstrophy flux $\Pi_Z$. 
    Parameters values are: 
    $m_f=6,M_f=240$, $\nu_{3}=10^{-18}$, $\tau^{-1}_1=2$, $F_0=1$}
  \label{fig:sp_e4}
\end{figure}
%%%%%%%%%%%%%%%%%%%%%%%%%%%%%%%%%%%%%%%%%%%%%%%%%%%%%%%%%%%%%%%%%%%%%%%%%%%%%%%%%%%%%%%%%%%%

Finally, Fig.\ref{fig:sp_e4} evinces the onset of a direct cascade
for $\varepsilon\,>\,3$. The enstrophy flux, 
\begin{eqnarray}
\label{numerics:enstrophyflux}
\Pi_{Z}(k,m_{f})=\int_{m_{f}}^{k}\frac{d^{2}q}{(2\,\pi)^{2}}\Re
\int d^{2}x\,e^{\imath \boldsymbol{q\cdot x}}
\prec\check{\omega}(-\boldsymbol{q},t)(v^{\beta}\de_{\beta} \omega)(\boldsymbol{x},t)\succ
\end{eqnarray}
is approximately constant and positive (indicating transfer towards 
smaller spatial scales) in the numerically resolved scaling range 
(see inset of Fig.\ref{fig:sp_e4}).
It should be emphasized that  numerical evidences of
\cite{NaOtAnGu00,TsOtAnGu05,BoCeMuVe02} uphold non-universal dependence
of the kinetic energy spectrum upon the Ekman friction in the direct cascade
regime. 
Since the analysis of such effects lays beyond the scopes of the present work
we replaced for $\eps\,>\,2$ the Ekman friction with an hypo-dissipative term 
$(-1)^{q+1}\tau_{q}^{-1}\de^{-2\,q}v$ which is expected to 
suppress the aforementioned non-universal corrections to the spectrum \cite{LiAv99}.

\section{Discussions and Conclusions}
\label{sec:discussion}

Our numerical experiments support without possible ambiguity the scenario
set by Kraichnan's phenomenological theory. The physically relevant order parameters
to describe the qualitative behavior of (\ref{model:ns}), (\ref{model:ps})
are the total energy (\ref{model:Einjection}) and enstrophy (\ref{model:Zinjection})
injections. Whenever they coherently act on small (large) scales an inverse
(direct) cascade is observed. In the intermediate case $2\,<\,\varepsilon\,<\,3$
no cascade takes place and a local balance scaling takes place. The renormalization
group scaling exponents are for $\varepsilon\,,h\,<\,2$ at most a sub-leading
correction to the inverse cascade scaling. It remains to be clarified
the origin of the discrepancy between the two theories. The renormalization group
analysis of \cite{Ho98,HnHoHoSe99,HoKaKo02} is very thorough and satisfies all
the self-consistency requirements that in the theory of critical phenomena are
known to produce correct scaling predictions. Thus any trivial explanation of 
the discrepancy can be safely ruled out. The indication of the existence of 
a constant-flux solution stems from the K\'arm\'an-Howarth-Monin equation. In the
renormalization group language the relations closest to (\ref{Kraichnan:KHM})
are those satisfied by the composite operator algebra which includes the energy 
dissipation operator (see e.g. section 2.2 of \cite{AdAnVa}). One may speculate
that non-locality of forcing plays a role different than in $3d$ in determining
the scaling dimensions of the elements of the operator algebra. It is however 
difficult to see how to consistently formalize this observation. 
One scenario that deserves to be further investigated is  in our opinion the following. 
Perturbative renormalization implies the
assumption that the infra-red stable fixed point governing the scaling regime emerges
from a bifurcation at marginality from a Gaussian fixed point. Such an assumption
is usually verified in critical phenomena but is not a necessary consequence of a
general non-perturbative theory. In particular, there are examples of field theories
where it is possible to give evidence that scaling is dominated by a 
fixed-point emerging at marginality from bifurcations from non-perturbative, non-Gaussian
fixed points. A concrete case is discussed in \cite{LiFi86,LiFi87} 
\footnote{We thank L.~Peliti for drawing our attention to 
this point and to refs. \cite{LiFi86,LiFi87}.}. Here a model of wetting transition
indicates a scenario which could apply also to $2d$-turbulence. The scaling predictions 
associated to an infra-red stable fixed point captured by perturbative renormalization 
group analysis is numerically seen to be dominated by those associated to a second fixed 
point not bifurcating from the Gaussian fixed point at marginality. The existence of this 
fixed point can be exhibited only by a non-perturbative construction of the renormalization 
group transformation. The price to pay is however the introduction of truncations of the
Wilson recursion scheme which cannot be a-priori fully justified. If we accept 
such point of view, the existence of the K\'arm\'an-Howarth-Monin equation should be 
interpreted as an a-priori indication of the existence of a non-perturbative 
fixed point.

\section*{Acknowledgements}

The authors are grateful to M.~Cencini, J.~Honkonen, A.~Kupiainen, L.~Peliti and M.~Vergassola
for numerous discussions and insightful comments. This work was supported by 
the center of excellence {\em ``Analysis and Dynamics''} of the Academy of Finland.

\appendix

\section{Asymptotics of the forcing correlation}
\label{ap:forcing}

Consider the scalar correlation
\begin{eqnarray}
  \label{ap:correlation}
  F\left(\boldsymbol{x},m,M\right)=F_{o}\int\frac{d^{d}p}{(2\,\pi)^{d}}\frac{e^{\imath \boldsymbol{p \cdot x}}}{p^{d+\eta}}
\,\chi\left(\frac{m^{2}}{p^{2}}\,,\frac{p^{2}}{M^{2}}\right)
\end{eqnarray}
with
\begin{eqnarray}
\label{ap:chi}
\chi(0,0)=1
\end{eqnarray}
In order to extricate the asymptotics in the range $m\,\ll\,p\,\ll\, M$ 
two cases should be distinguished depending upon the sign of $\eta$. 
\begin{itemize}
\item If $\eta\,<\,0$ the integral is infra-red and ultra-violet 
convergent at finite point separations in the absence of 
cut-offs:
\begin{eqnarray}
\label{ap:neg}
F\left(\boldsymbol{x},m,M\right)\simeq F_{o}\int \frac{d^{d}p}{(2\,\pi)^{d}} 
\frac{e^{\imath \boldsymbol{p \cdot x}}}{p^{d+\eta}}\,=\,
\frac{\Omega_{d}}{(2\,\pi)^{d}}
\frac{x^{\eta}\,\Gamma\left(\frac{d}{2}\right)\,\Gamma\left(-\frac{\eta}{2}\right)}{
2^{1+\eta}\,\Gamma\left(\frac{d+\eta}{2}\right)}
\end{eqnarray}
with $\Omega_{d}$ the solid angle in $d$-dimensions.
\item  If $\eta\,>\,0$ (\ref{ap:correlation}) is convergent only if the 
infra-red cut-off is retained:
\begin{eqnarray}
\label{ap:pos}
  F\left(\boldsymbol{x},m,M\right)\simeq F\left(\boldsymbol{x},m,\infty\right) 
=F_{o}\int\frac{d^{d}p}{(2\,\pi)^{d}}
  \frac{e^{\imath \boldsymbol{p \cdot x}}}{p^{d+\eta}}
\,\chi\left(\frac{m^{2}}{p^{2}}\,,0\right)
\end{eqnarray}
The integral can be estimated by inverting its Mellin representation with 
help of Cauchy theorem (see e.g. \cite{KuMG07})
\begin{eqnarray}
\label{ap:Mellin}
%\hspace{-1.8cm}
\hspace{-0.8cm}
F(\boldsymbol{x},m,\infty)=%\frac{\Omega_{d}}{(2\,\pi)^{d}}
F_{o}\int_{\Re\zeta-\imath\,\infty}^{\Re\zeta+\imath\,\infty}\hspace{-0.2cm}
\frac{d\zeta}{(2\,\pi\,\imath)}\,x^{\eta}(m\,x)^{2\,\zeta}\,\phi(\zeta)
%\frac{ x^{\eta}(m\,x)^{2\,\zeta} \tilde{\chi}(\zeta)\,
%\Gamma\left(\frac{d}{2}\right)\,\Gamma\left(-\frac{2\,\zeta+\eta}{2}\right)}{
%2^{1+2\,\zeta+\eta}\,\Gamma\left(\frac{d+2\,\zeta+\eta}{2}\right)}\,,
\,,\hspace{0.9cm}
\Re\,\zeta\,<-\frac{\eta}{2}
\end{eqnarray}
The holomorphic function
\begin{eqnarray}
\label{ap:Mellinampl}
%\tilde{\chi}(\zeta):=
\phi(\zeta):=\frac{\Omega_{d}}{(2\,\pi)^{d}}
\frac{\Gamma\left(\frac{d}{2}\right)\,\Gamma\left(-\frac{2\,\zeta+\eta}{2}\right)}{
2^{1+2\,\zeta+\eta}\,\Gamma\left(\frac{d+2\,\zeta+\eta}{2}\right)}
\int_{0}^{\infty}\frac{dw}{w}\frac{\chi(w^{2},0)}{w^{\zeta}}
\end{eqnarray}
is by hypothesis analytic at least in a stripe for $\Re \zeta\,<\,-\eta/2$. 
Furthermore, by (\ref{ap:chi}), the integral in (\ref{ap:Mellinampl}) 
generates a simple pole for $\zeta$ equal zero.
Cauchy theorem yields for $m\,x\,\ll\,1$ the asymptotics
\begin{eqnarray}
\label{ap:Mellinres}
%\hspace{-2.5cm}
F(\boldsymbol{x},m,\infty)=
%\frac{\Omega_{d}}{(2\,\pi)^{d}}m^{-\eta}\left\{
%F_{o}\tilde{\chi}\left(-\frac{\eta}{2}\right)+
%\frac{x^{\eta}\,\tilde{F}_{o}\,\Gamma\left(\frac{d}{2}\right)\,
%\Gamma\left(-\frac{\eta}{2}\right)}{2^{1+\eta}\,\Gamma\left(\frac{d+\eta}{2}\right)}-
%\frac{(m\,x)^{2}\,F_{o}\tilde{\chi}\left(1-\frac{\eta}{2}\right)}{d}
%+\dots\right\}
F_{o}\,m^{-\eta}\left\{\bar{\phi}_{-\frac{\eta}{2}}+
(m\, x)^{\eta}\,\bar{\phi}_{0}+(m\,x)^{2}\,\bar{\phi}_{1-\frac{\eta}{2}}
+\dots\right\}
\end{eqnarray}
having used the notation
\begin{eqnarray}
\label{ap:residue}
\bar{\phi}_{a}:=-\lim_{\zeta\,\uparrow\,a}(\zeta-a)\,\phi(\zeta)
\end{eqnarray}
\end{itemize}

\section{Large scale zero-mode and power-law forcing in the Kraichnan model of advection of a concentration field}
\label{ap:KM}

We refer the readers for definition and details on the large scale decay
properties of the  Kraichnan model to \cite{FaFo05,CeSe05,CeSe06,MaMG09}. 
%For \emph{non-vanishing} forcing charge the decay properties are associated 
%to the large scale zero mode of the model
%\begin{eqnarray}
%\prec\theta(\boldsymbol{x},t)\theta(0,t)\succ
%\overset{x\uparrow \infty}{\sim}\frac{1}{x^{d-2+\xi}}
%\end{eqnarray}
%yielding the prediction for the spectrum
The energy spectrum of the Kraichnan model is exactly known. In $2\,d$
it takes, modulo irrelevant constant factors, for isotropic forcing
the form
%\begin{eqnarray}
%E_{[\theta]}(p)= \int_{0}^{\infty}d\rho\,\frac{B_{1}(p\,\rho)}{\kappa+D\,\rho^{\xi}}
%\int\frac{d^{2}q}{(2\,\pi)^{2}}\frac{\rho\,B_{1}(q\,\rho)}{q}\check{G}(q)
%\label{ap:spectrum}
%\end{eqnarray}
\begin{eqnarray}
E_{[\theta]}(p)= \int_{0}^{\infty}\frac{d\rho}{\rho}\,\frac{\rho\,B_{1}(p\,\rho)}{\kappa+D\,\rho^{\xi}}
\int_{0}^{\rho}\frac{d\sigma}{\sigma}\,\sigma^{2}\,G(\sigma,m_{g},M_{g})
\label{ap:spectrum}
\end{eqnarray}
with $B_{\eta}(x)$ the Bessel function of order $\eta$, $D$ the eddy 
diffusivity of the advecting velocity field and $G$ given by (\ref{model:G}).  
The H\"older exponent $\xi$ is a free parameter in the model. 
Turbulent advection corresponds to $\xi$ equal $4/3$. From (\ref{ap:spectrum}) it
is straightforward to check that
\begin{eqnarray}
\label{ap:asymptotics}
%\hspace{-2.4cm}
\hspace{-1.0cm}
E_{[\theta]}(p)\sim \left\{
\begin{array}{ll}
p\,\int_{0}^{\infty}\frac{d\rho}{\rho}\, 
\frac{\rho^{2}}{\kappa+D\,\rho^{\xi}}\frac{\tilde{G}_{o}}{\rho^{2-2\,h}}=
\left(\frac{\kappa}{D}\right)^{\frac{2\,h}{\xi}}
\frac{\tilde{G}_{o}\,\pi\,p}{\kappa\,\xi\,\sin(\frac{2\, h\, \pi}{\xi})}\,, 
\hspace{0.4cm}&\hspace{0.4cm} 
h\,<\,\frac{\xi}{2}
\\[0.5cm]
p^{1+\xi-2\,h}\int_{0}^{\infty}\frac{d\,\rho}{\rho}\, 
\frac{\tilde{G}_{o}\,B_{1}(\rho)}{D\,\rho^{1+\xi-2\,h}}
=\frac{\tilde{G}_{o}\,p^{1+\xi-2\,h}\,\Gamma\left(h-\frac{\xi}{2}\right)}
{2^{2+\xi-2\,h}\,D\,\Gamma\left(2-h+\frac{\xi}{2}\right)}\,,
\hspace{0.4cm}&\hspace{0.4cm} h\,>\,\frac{\xi}{2}%\,<h\,<\,2
%\\[0.5cm]
%\check{G}(0)\,p^{\xi-1}\int_{0}^{\infty}\frac{d\,\rho}{\rho}\, 
%\frac{B_{1}(\rho)}{D\,\rho^{\xi}}
%\hspace{0.4cm}&\hspace{0.4cm} \frac{\xi}{2}\,<h\,<\,2 \hspace{0.2cm}\& \hspace{0.2cm}
%\check{G}(0)\,>\,0
\end{array}
\right.
\end{eqnarray}
$\tilde{G}_{o}$ is a dimensional constant, the value of which is 
irrelevant for the present considerations.
Setting $\xi=4/3$ the results of subsection~(\ref{sec:inverse})
of the main text are recovered. The direct cascade results (\ref{direct:spectra}) 
are recovered instead by setting $\xi=2$.
The asymptotics (\ref{ap:asymptotics})
hold under the assumption of infinite integral scale $m_{g}^{-1}$ of
of the spatial forcing correlation. In the language of \cite{CeSe05,CeSe06}
this means that the "charge" $\check{G}(\boldsymbol{0},0,M_{g})$ is 
\emph{vanishing}. 
For $h>2$ the forcing becomes infra-red dominated. At scales $m_{g}x\, \ll\, 1$

\begin{eqnarray}
%\label{}
E_{[\theta]}(p)\sim G(0,m_{g},\infty)\,p^{\xi-3}\int_{0}^{1}\frac{d\,\rho}{\rho}\, 
\,\rho^{3-\xi}\frac{B_{1}(\rho)}{D}
\end{eqnarray}
whilst in the opposite range $m_{g}x\, \gg\, 1$, for $\check{G}(\boldsymbol{0},m_{g},\infty)>0$

\begin{eqnarray}
%\label{}
E_{[\theta]}(p)\sim \check{G}(0,m_{g},\infty)\,p^{\xi-1}\int_{0}^{\infty}\frac{d\,\rho}{\rho}\, 
\,\rho^{1-\xi}\frac{B_{1}(\rho)}{D}
\end{eqnarray}

\end{document}